%% file: main.tex
\documentclass[submission,copyright,creativecommons]{eptcs}
\providecommand{\event}{FESCA 2016} 
\usepackage{breakurl}             

\usepackage{amssymb}
\usepackage{graphicx}

\usepackage{float}
\usepackage{verbatim}
\usepackage{xcolor}
\usepackage{multirow}
\usepackage{lscape}
\usepackage{caption}

\usepackage{wrapfig}
\usepackage{amssymb}

\usepackage{listings}

\definecolor{dkgreen}{rgb}{0,0.6,0}
\definecolor{gray}{rgb}{0.5,0.5,0.5}
\definecolor{mauve}{rgb}{0.58,0,0.82}
\definecolor{light-gray}{gray}{0.25}
 
\usepackage{listings}
 \renewenvironment{abstract}  
 {\noindent
 {\medskip}
  \small
  \list{}{%
    \setlength{\leftmargin}{9mm}
    \setlength{\rightmargin}{\leftmargin}%
  }%
  \item\relax}
 {\endlist}

\usepackage{ragged2e}
\usepackage{url}
\usepackage{color}
\usepackage{adjustbox}

\usepackage[textsize=footnotesize,shadow]{todonotes}

\usepackage{url}
\urldef{\mailsa}\path|{gordon.pace, christian.colombo, luke.chircop}@um.edu.mt|

\begin{document}

\input{title}

\input{abstract}
\input{introduction}

\input{monitoring}

\input{android}

\input{specification-language}

\input{case-study}
\input{related-work}
\input{conclusions}

\bibliographystyle{eptcs}
\bibliography{ref}

\end{document}

%% file: title.tex
\title{Device-Centric Monitoring for Mobile Device Management\thanks{The research work disclosed in this publication is partially funded by the \textit{Master it!} Scholarship Scheme (MALTA).}}

\author{Luke Chircop \quad\quad Christian Colombo \quad\quad Gordon J. Pace
\institute{Department of Computer Science}
\institute{University of Malta}
\email{\{luke.chircop | christian.colombo | gordon.pace\}@um.edu.mt}
}

\def\titlerunning{Device-Centric Monitoring for Mobile Device Management}
\def\authorrunning{L. Chircop, C. Colombo and G. J. Pace}

\maketitle

%% file: abstract.tex
\begin{abstract}
 The ubiquity of computing devices has led to an increased need to ensure not only that the applications deployed on them are correct with respect to their specifications, but also that the devices are used in an appropriate manner, especially in situations where the device is provided by a party other than the actual user. Much work which has been done on runtime verification for mobile devices and operating systems is mostly application-centric, resulting in global, device-centric properties
  (e.g. the user may not send more than 100 messages per day across all applications) being difficult or impossible to verify. In this paper we present a device-centric approach to runtime verify the device behaviour against a device policy with the different applications acting as independent components contributing to the overall behaviour of the device. We also present an implementation for Android devices, and evaluate it on a number of device-centric policies, reporting the empirical results obtained.

\end{abstract}

%% file: introduction.tex
\section{Introduction}


Technology is constantly becoming more pervasive in our daily lives, especially with the case of portable electronic devices which are being used in a variety of areas including entertainment, work, study, health, finance, etc. 
Whilst this brings in new opportunities for more applications supporting our personal life, the need to ensure that the devices work, and are used as expected, and that data does not leak from one application to another becomes more important. Current operating systems such as Android, iOS, etc, are well equipped to manage such risks by strictly keeping the applications separate and only with the user's permission can data flow from one application to another. 
However, when the device user is not the device owner, the existing mechanisms fall short of providing the necessary safeguards. 

For example, consider a company handling goods distribution, and which gives its employees portable electronic devices in order to help them promote new products, register sales orders, produce invoices, etc. In doing so, the distribution company would have to store commercially sensitive data on the device without having the ability to control when, how and by whom it is accessed. 

Apart from the lack of control over sensitive data, the company might also become concerned with how its employees use the device. These include the time they spend on social media and the installation and use of third-party applications. Similar concerns would typically also be significant for parents providing mobile devices to their children. 
 
Tackling such issues might force a device owner to commission a custom solution to limit the usability of the device.  
However, this approach can be costly (e.g.\ not affordable to parents) and inflexible, since modifying the introduced limitations would typically require the involvement of technical experts. 
An alternative would be to have off-the-shelf tools which are easily configurable by the device owner. 
A number of such solutions \cite{runtimeVerification,runtimeVerification2,runtimeVerification3} employ runtime verification techniques to automatically generate monitors from device-owner-specified rules --- referred in the sequel as properties.  

Although there are tools that are capable of monitoring and verifying user behaviour, they suffer from a significant limitation, namely that they are application-centric rather than device-centric, i.e.\ they do not support properties spanning multiple applications such as: No more than 100 messages can be sent by the device per hour. 
If existing application-centric tools are used to monitor such property, a device user could easily avoid having the property violated by installing multiple applications capable of sending messages and sending less than a hundred messages per application. 

In the rest of the paper, we describe our solution to provide device-centric monitoring. More specifically,  we start by discussing the design of such an approach in Section \ref{design}. Subsequently, we describe how the design has been instantiated for Android (Section \ref{androidInstantiation}), introduce the Specifications used to accept device-centric properties (Section \ref{sec:specificationLang}), and report the empirical results of its application to a case study in Section \ref{caseStudy}. Finally we discuss related work (Section \ref{sec:relatedWork}) and conclude the paper by listing a number of possible future ways of extending the work (Section \ref{sec:conclusions}).

%% file: monitoring.tex
\section{Device-Centric Monitoring}
\label{sec:monitoring}
\label{design}

Portable devices nowadays offer the ability to use a multitude of services ranging from traditional features like message sending and receiving capabilities, to the use of GPS, NFC and Wi-Fi functionality. Furthermore, these abilities are typically accessed through the use of external applications such as browsers, games and maps, which are typically run separately and with limited control across applications in order to address security and privacy concerns. When functionality is to be limited for a single application on the device e.g. `The stock control app may send messages to persons outside the domain of the company', or `The stock control app may not send more than 100 messages in a single day', standard techniques for restricting system behaviour can be applied --- from statically applied app policies which do not allow the application to be installed in the first place, to dynamic approaches which allow only applications with instrumented monitors to ensure correct behaviour. The challenge arises when such policies span across multiple application. For instance, a policy which states that `The device may not be used to send more than 100 messages per day' or `Text copied from the stock control app may not be pasted into a messaging app' cannot be checked by looking at the applications individually, but necessarily require a way of centralising or sharing information across applications.    




Due to the isolation in which mobile applications operate, device-wide monitoring  effectively presents the same challenges as globally monitoring component-based distributed systems. In the literature, one finds two distinct approaches to addressing this issue: orchestration and choreography \cite{dist}. Orchestration-based monitoring sets up a central monitor that has access to the relevant events in the components and performs all the necessary checks to identify property violations e.g. \cite{orchestration}. In contrast, choreographed approaches decompose monitoring functionality into parts which are integrated across the components, with any required monitor dependencies across components being handled using direct communication across the relevant components e.g. \cite{choreo1,choreo2,choreo4}. 

Since a number of tools enabling monitor instrumentation into individual applications already exist \cite{RVDroid,WeaveDroid,xu2012aurasium} and there is a body of work on choreographed organisation of properties \cite{dist,choreo2}, such an approach would enable device-wide monitoring provided inter-monitor communication is possible. This communication would be required, for instance when monitoring that `The stock control app may not send more than 100 messages in a single day', since the total number of messages would have to be accessible to the local monitors instrumented in each app. 
However, for security reasons, mobile device operating systems allow communication across applications only through user-granted permissions. This makes the choreography approach infeasible unless major changes are done to the security framework of the operating system in question. Even if such communication was possible, there still remain a number of issues such as where to store data which belongs to the property (e.g. the message count): Would it be replicated on all message-sending applications or on one of them which is selected to act as property manager? Another issue deals with permissions: Monitors might require special privileges, namely to obtain sensitive device information such as the battery status, and to disallow the user from being able to interfere with monitoring processes. If monitors are incorporated within applications (which run in the user space) such privileges would be tricky to provide.  


Furthermore, in choreographed systems, properties which require some form of shared memory (such as the message count in the past day) would require such information to be stored in a new separate process (in effect this process can be seen as an orchestrator) or by one of the existing processes, in which case closing down the master application which keeps the count would interfere with the monitoring.

An alternative monitor organisation approach used for component-based system monitoring is orchestration. An orchestration-based approach, having a centrally-located orchestrator (as opposed to within applications) on the device, can address the issues encountered with a choreography:

\begin{description}
  \item[Communication:] Having a central orchestrator avoids the need to have each monitor communicate to its peers. While this does not solve the problem \emph{per se}, it simplifies it.
  \item[Shared memory across applications:] The centralised monitor itself can be used to store and manipulate state required to be shared across applications (e.g. count of messages sent). 
  \item[Access to device information:] Device information (e.g. battery status) can only be accessed by applications if they have the right permissions. Having a centralised monitor avoids having to update these permissions and, in the process, also ensures information is not provided unnecessarily to applications. Similarly, information such as whether an application has access to files written by another application will also be visible from a central viewpoint.
  \item[Setting of device attributes] In a similar manner, a central monitor allows the change of application or system attributes without requiring giving explicit permissions to individual applications. For instance, setting display brightness, or disabling access to certain services, can be performed by the central monitor without additional permissions.
  \item[Protection from user interference:] Having monitors centrally located at system level and not alongside or as part of the applications in user space, also ensures they can be protected from user tampering (e.g. users killing a monitoring process in order to give themselves unconstrained control of the device). 
\end{description}

While orchestration solves several problems, it has limitations of its own, in particular that the central orchestrator might not have enough visibility and control of internal application behaviour, e.g. having access to Skype message send requests and the ability to block such requests from going through. Therefore, we propose a hybrid approach consisting of a central orchestrator and monitoring fragments strategically placed within applications to transmit the required information to the central node.


In what follows, we explain how this design has been implemented in the context of the Android operating system.

%% file: android.tex
\section{An Android Instantiation}
\label{sec:androidInstantiation}
\label{androidInstantiation}

Identifying a way of implementing the design with a mix of orchestration-choreography is not straightforward. The first concern is how to instrument the monitoring components, and second is how to enable them to communicate with each other. The rest of the section expands on these issues. 

\subsection{Instrumenting monitors}

\begin{figure}[t]
	\begin{center}
		\includegraphics[width=0.65\textwidth]{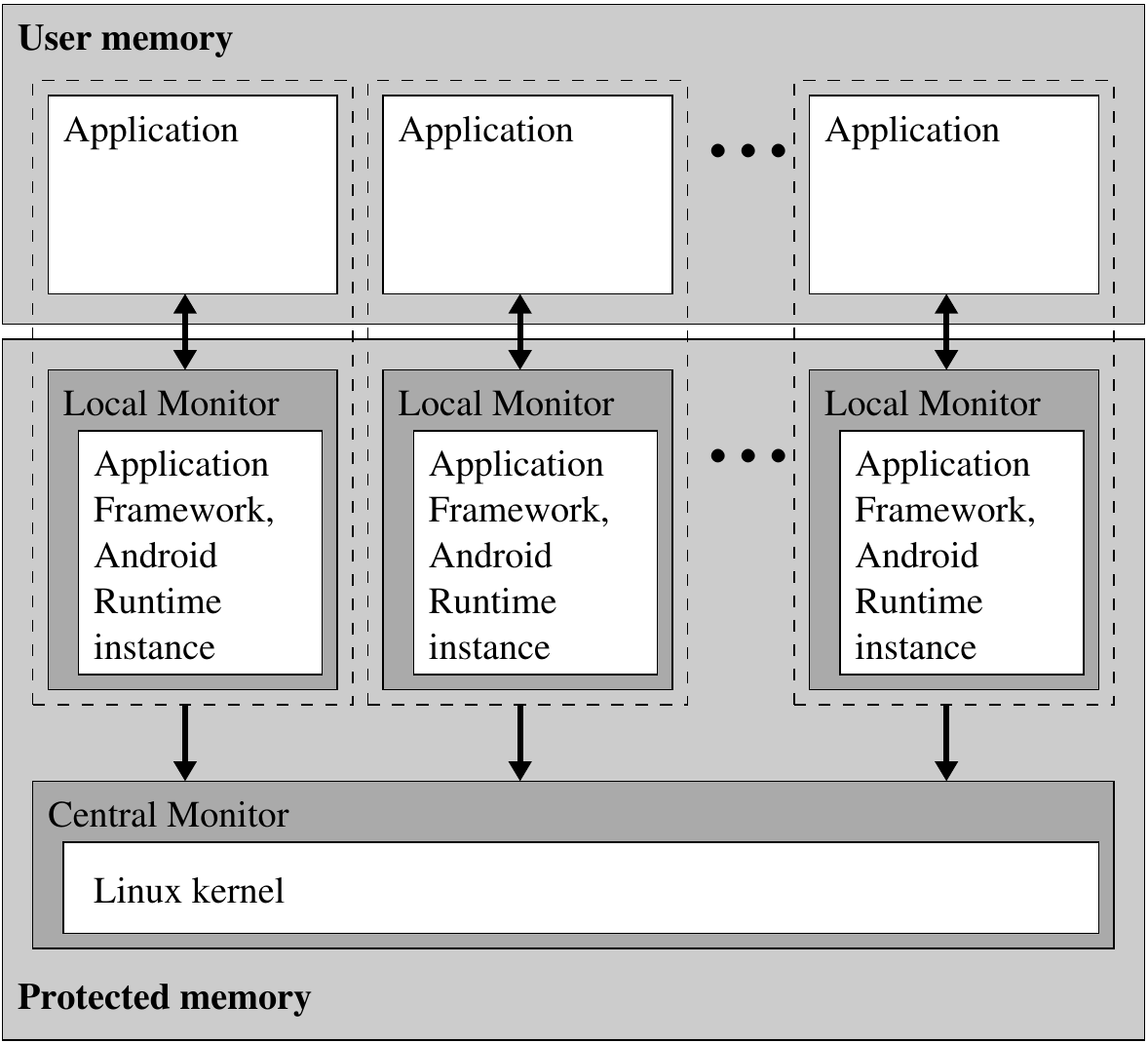}
	\end{center}    
  \caption{Runtime verification architecture for the Android operating system}
	\label{fig:rvarchitecture}
\end{figure}

On an Android device, the central OS kernel loads a separate instance of the Dalvik virtual machine  and runtime environment for each application, which includes an application framework together with an Android runtime environment. Calls from the application to the environment thus all go through this layer before being passed on the Android kernel through system calls if the need arises. Since our intention is to be able to monitor the interaction between the applications and the device, an ideal point of instrumentation for the monitoring is either within the virtual machine instances or the kernel itself. Either way, user tampering of the monitoring processes is ruled out since the user does not have write-access to either. 
As illustrated in Figure~\ref{fig:rvarchitecture}, we have modified the system such that all instances of the virtual machine created by the kernel now include a listener which has access to all calls made from the applications to the virtual machine. Effectively, each instance of an app, corresponding to a vertical column in the Figure~\ref{fig:rvarchitecture}, has a layer of monitoring logic which is local to the app, yet residing in protected memory. 

However, instrumenting applications individually is not enough as this would not support monitoring of device-wide property monitoring. Therefore, the next step is to instrument the central orchestrator at an appropriate point in the system. The solution we have developed is to instrument a central monitor at the kernel level which allows the system to keep track of global property states. This hybrid centralised and decentralised monitoring approach allows for device-centric monitoring in a manner which is application agnostic. Again, the central monitoring component is depicted in Figure~\ref{fig:rvarchitecture}.

\subsection{Communication}


As already discussed, communication between applications may be restricted since it is only allowed if given the necessary permissions. This posed a challenge for our implementation since there was a high probability of having monitors running with applications that that not have such permissions. This makes the use of sockets, for instance, to communicate between application and global monitors difficult and not always possible. Modifying the system so that it provides the required permissions to all running applications was not seen as an option since this would have introduced a potential vulnerability that could easily be exploited by malicious applications. 

System calls are typically used by applications to make use of kernel functionality, implicitly a synchronisation and communication point from application- to kernel-side code. In our framework, we exploit this ability to enable communication between the localised and the central monitor --- adding new system calls which are used to act as synchronisation points and data passing channels to the central monitor. To communicate with the global monitor, the application-side monitor thus calls the corresponding system call, passing all the necessary information through arguments and waiting for a reply encoding the result in the system call return value. 

%
%

The main limitation of this approach is that, since the central monitor is effectively a group of system calls, unsolicited communication from the central monitor to a particular application not possible. Instead our approach relies on having the application side monitor initiate the communication and then having the central monitor relay back information or action instructions for the application side monitor to process and execute.

\subsection{Discussion}

Having a hybrid of centralised and decentralised monitoring helped address a number of issues discussed in Section \ref{sec:monitoring}. Protection from users and applications is one of the main concerns when introducing runtime monitors that control their behaviour. One would need to ensure that the monitors cannot be affected or tampered with by unauthorised users and applications. To this end, our framework has achieved this by design since the centralised monitor is embedded inside the Android kernel which cannot be modified by third-party applications making it temper proof. Furthermore, the decentralised monitors are also temper proof since they are instrumented inside the Android middleware which again cannot be modified by users or third-party applications. 

It is also important to have the monitor's state to be protected from unauthorised tempering. 
Our implementation \footnote{The implementation is available at: \url{https://github.com/lukechircop007/DeviceCentricMonitoring.git}} has achieved this by having the state of the centralised monitor states located at kernel side memory and the decentralised monitor states located inside the protected memory both of which cannot be tempered with by users and applications. 
Given the nature of the framework's implementation and versatility of the language specification, access to device information and setting of device attributes can also be easily achieved through the use of functionality that the OS provides by default. 
The only possibility of having the instrumentation introduced by our framework tempered with or bypassed is by having a fresh OS installed on the device. 

%% file: specification-language.tex
\section{Device-Centric Monitoring Specifications} 
\label{sec:specificationLang}

Due to the shared features between device-centric monitoring and component-based system monitoring, in that we have different applications and a central monitor coordinated together, we have opted to use the runtime verification tool polyRV \cite{polylarva} as our starting point. The tool polyRV has been developed to enable the specification of monitors which span across multiple components, possibly made up of different technologies. For this reason, the tool allows subparts of the specification to be tagged according to the component they deal with, including parts of the monitor which do not belong to any particular system component (e.g. maintaining a global counter across components) and is tagged as belonging to the central monitoring component. 

Following this approach, our specification language has parts of the specification tagged as either \emph{GlobalSide} or \emph{ApplicationSide}. Since, in our case, we do not access the internals of the applications (i.e. components), application-side specifications are automatically replicated for all applications, making it is unnecessary to specify which application the specification part refers to. If one desires to act in a more application specific manner, one can still access application name and identifier. 

A device-centric specification consists of a number of rules, each being a triple: $e \mid c \mapsto a$, where $e$ is an event, $c$ is a condition and $a$ an action. Events are observable points in the behaviour of the applications (e.g. the moment a message sending attempt is being made by an application, or the moment the application accesses a website). The condition is any expression used to check predicates over accessible state on the application-side, kernel-side, success (or otherwise) of system calls and monitoring state (e.g. a counter variable which is created solely for the monitoring requirements). Finally, actions are fragments of executable code which can be executed to update any such writeable state, whether on the application, kernel or the monitoring state. All three parts of the rules can be tagged to be application or kernel side. 

The semantics of these rules corresponds to those of the polyRV specification language \cite{polylarva}. Informally, upon the occurrence of an event, each rule which is triggered by that event is considered --- its condition is evaluated, and if it evaluates to true, then the action is executed. 

Consider, for instance, specifying the property limiting messages to 100 per day:
{\footnotesize 
\begin{lstlisting}[numbers=none, label=polyRV3]   
Events {	
   ApplicationSide { 
      sendMessageBefore() = {
         SmsManager *.sendTextMessage(...)      
      } 
      sendMessageAfter(boolean sent) = {
         after SmsManager *.sendTextMessage(...)
      } uponReturning(sent)
   }
} 
Conditions {
   GlobalSide { maximumQuotaReached = ... }
   ApplicationSide { successful = {sent == true} }
} 
Actions {
   ApplicationSide { blockRequest = {applicationSide {return;}}}
   GlobalSide { incrementMessageCount = {globalSide{...}}}
}
Rules {
   sendMessageBefore() |  maximumQuotaReached()               -> blockRequest();
   sendMessageAfter()  | !maximumQuotaReached() && successful -> incrementMessageCount();
}
\end{lstlisting}
}


%
%
%
%


In the script above, \textit{sendMessageBefore} and \textit{sendMessageAfter} are application-side events which correspond to the moments before and after a message send request; \textit{successful} represents an application side condition checking that the message request has been successfully executed; whilst \textit{maximumQuotaReached} represents a global side condition that will check whether the global maximum has been reached. Finally \textit{blockRequest} and \textit{incrementMessageCount} are the actions that trigger once their corresponding conditions are satisfied, effectively allowing us to enforce. 

Since the events we capture  (\textit{sendMessageBefore} and \textit{sendMessageAfter}) happen on the application-side, they are tagged to be \textit{ApplicationSide}. Similarly, whether the message has been successfully sent uses the value returned to the application and is thus also tagged \textit{ApplicationSide}. On the other hand, we keep a central monitoring state to check the number of messages sent, meaning that access to this state (\textit{maximumQuotaReached}, \textit{incrementMessageCount}) are specified to be \textit{GlobalSide} . Finally, blocking further messages from being sent has to occur directly on the application side\footnote{Another approach would have been to extend the (global) monitoring state to store whether messages are to be blocked, and then bypass message sending at the kernel-side whenever this flag is set.}.

\begin{figure}[t]
	\begin{center}
		\includegraphics[width=0.85\textwidth]{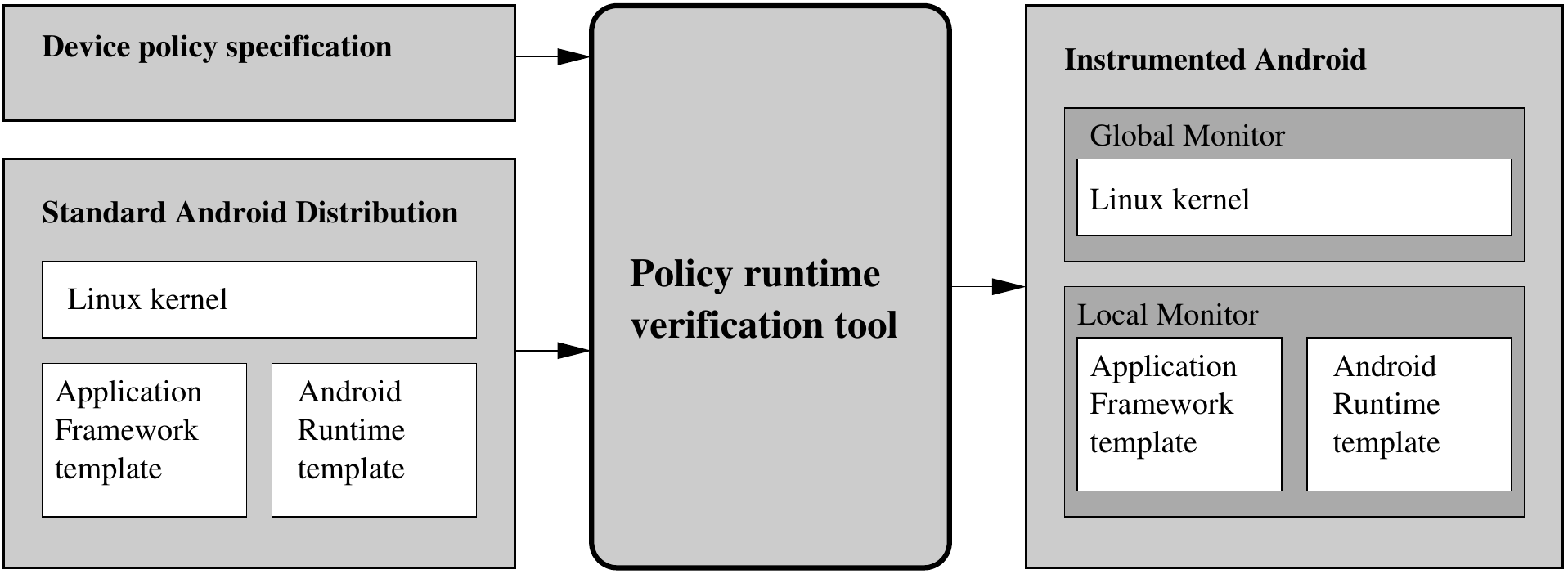}
	\end{center}    
  \caption{Runtime policy verification tool for the Android operating system}
	\label{fig:toolflow}
\end{figure}

A runtime monitoring proof-of-concept tool has been built for the specification language described, which instruments a given device policy into an Android distribution as local and global monitors as shown in Figure~\ref{fig:toolflow}. The local monitors are instrumented in the Android Runtime and the application framework templates, which are instantiated by the OS for each application execution. The core of the OS is also instrumented with code to manage the global parts of the monitor. Therefore, given a policy specification and an Android distribution\footnote{Currently, the tool works for a particular distribution, in which we manually inserted hooks to enable us to instrument the policies more easily. It could, however, be extended to work on general distributions (with some constraints).}, we can produce a policy-instrumented Android distribution which can be compiled and deployed onto a device, guaranteeing that the policy is adhered to, and cannot be bypassed unless the device is reinstalled with a policyless instance of Android. 

This tool has been applied on two case studies which are discussed in the following section. 

%% file: case-study.tex
\section{Case Study}
\label{caseStudy}

The approach presented in this paper has been applied to two case studies: (i) a family setting where parents want to control children's use of mobile devices, and (ii) a distribution company which provides mobile devices to its employees, but would like to enforce a usage policy to control their use.
 
\begin{description}
\item[Parental control (PC)]
A 2014 study \cite{childStudy} has identified parents buy or allow their children to use mobile devices for educational purposes but children mainly use the them for entertainment purposes. Ensuring that mobile devices are used for the purpose intended by the parent as well as ensuring the safety of the content would typically top the list of the parents' concerns. 
Parents might also be concerned with how much their children spend on phone calls and messages. 
The study has also shown that in most cases, children use mobile devices for more than one hour per sitting potentially causing health issues. 

\item[Distribution company (DC)]
The second case study is that of a coffee distributor that imports various coffee-related products and sells them to hotels, restaurants, groceries, stores, etc. In order to do so, the company has its employees visiting every customer in person to promote new and existing products that are for sale, place orders and deliver them whilst providing the necessary receipts. In order to facilitate the process, the company entrusts its employees with mobile devices that contain all the required information such as customer (account number, orders,  business details, etc) and product details (name, price, quality, stock, etc.).  The company is thus concerned that the employees, who have access to all the sensitive data could potentially access them for their personal gain without the company's permission. Another issue is that of stolen and lost mobile devices, potentially compromising sensitive data to unauthorized people outside the company. 

Given the fact that mobile devices allow their users (in this case employees) to install and use a variety of applications, companies might also be concerned that their employees would install malicious applications with or without their intention, potentially jeopardizing the device and its contents.  The company is also concerned with the amount of time their employees spend on particular applications (such as social media applications and games) which could result in poor performance at work, hurting company profits. Even if an application is allowed, the company might thus also want to not disclose the current location that its employee is currently in. Browsing can also prove to be of a concern to a company since there might be some URL requests that could be inappropriate and might taint the company's reputation. Similar to the family scenario, the company may also be concerned with how much their employees spend on calls or messages. 
\end{description}

A number of properties were found to be shared between the two scenarios so we present them in a single list classified according to property type and tagged with \emph{PC} and \emph{DC} depending on relevant scenario(s). Each property is given a label for easy referencing throughout the rest of the paper.

\begin{description}

\item[Prohibition]
{\ }
	\begin{description}
	\item[PhoneExtBlk (DC)] 
	Phone calls to foreign countries cannot be performed.
    \item[URLBlkReq (PC, DC)] 
    URLs containing inappropriate content must be blocked. 
    \item[WiFiLmt (DC)]
    During work hours, the Wi-Fi cannot be used or turned on. 
    \item[FileAccessLmt (PC, DC)]
    Access to particular folders should be restricted unless performed by authorized applications/users.  
	\end{description}
	
\item[Time limitation] 
{\ }
	\begin{description}
	\item[PhoneTimeLim (PC, DC)]
	Only four hours per day can be spent on phone calls.
	\item[GamePlayLmt (PC)] 
	No more than three hours per day can be spent playing games.
	\item[OneHrPerSittingOnly (PC)]
	The mobile device cannot be used for more than one hour per sitting and if surpassed, at least 15 minutes need to pass before the device can be used again
   \item[MsgTimeLmt (DC)]
   Message request cannot be performed before one minute has passed from the previous request. 
	\end{description}
	\pagebreak
\item[Time and count limitation]
{\ }
	\begin{description}
	\item[MsgCntLmt (PC)]
	No more than 500 messages can be sent per day.
    \item[MsgCntLmtHr (PC)]
    No more than 100 messages can be sent per hour. 
	\end{description}
\end{description}

All of the above properties have been specified using our specification language and transformed into runtime monitors. The process of implementing the properties was straightforward provided that the hooks into the relevant events required by the property were already in place. In order to extend this set, involved identifying API calls provided by the OS and annotating the Android distribution ready to be instrumented by our tool. The properties varied in complexity e.g. expressing \textit{WiFiLmt} was straightforward, requiring only a few rules and lines of code, while expressing \textit{FileAccessLmt} required much more effort since it required the maintenance of a access list to be built and maintained by the rules. 

Consider the \textit{WiFiLmt} property specification below: 

{\linespread{1}{\tiny \scalebox{.2}{supertiny}
\begin{lstlisting}[caption=Partial script representing WiFiLmt property, label=propp1]  
   events {	
      onWifiEnable() = {
         WifiManager *.setWifiEnabled(boolean enabled)
      }
   }	
   conditions {
      areEnabledRequests = {applicationSide{enabled == true}}	
      isTooEarly = {
      	applicationSide{
            Calendar c = Calendar.getInstance();
            int hour;
            String AM_PM;
            hour = c.get(Calendar.HOUR_OF_DAY);
            int ds = c.get(Calendar.AM_PM);
            if (ds==0) AM_PM="am"; else AM_PM="pm";
      		return (hour > 1 && AM_PM . equals (" pm "))
      	}
      }
   }
   actions {
      blockWifiRequest = {
         applicationSide{
               setWifiEnabled(false);
            }
      }
   }
   rules {
      blockWifiAfterHours = 
         onWifiEnable() \ areEnabledRequests && isTooEarly -> blockWifiRequest
   }
\end{lstlisting}
}}

The single rule catches the \textit{setWifiEnabled} API call inside the \textit{WifiManager} class as an event, retrieves the current time and determines accordingly if the WiFi can be turned on. 


\subsection{Empirical Results}

As with all runtime verification overheads induced by the monitors are a concern. For this reason, a series of tests were carried out measuring the performance and memory overheads that are introduced by device-centric properties varying in complexity. To be able to do so, the ``San Francisco'' \cite{mobile} ARMv6-based CPU running (CyanogenMod \cite{cyano})Android 4.3.1, kernel version 2.6.35.7 and 420MB RAM was used. 


\subsubsection{Measuring Performance Overheads}

The first experiment consisted of measuring the performance overheads that the monitoring framework introduces for the various properties. Since our monitors are event-triggered, no extra runtime overheads are introduced to unrelated processes or services. Therefore, we present the results for each instrumented event, comparing the duration of the event with and without the monitors. The measurements for each monitored and unmonitored event was taken a number of times and averaged to minimise variations from other sources such as background OS processes.

To obtain the base time for each event, we started off by measuring their execution times on a freshly installed Android operating system. Subsequently, we proceeded to instrument the monitors inside the Android operating system. Executing the same sequence of actions on the same applications, the time that the aforementioned events took to service the requests was measured. Given the uninstrumented and instrumented execution values, the total performance overheads introduced by the approach were calculated. The results can be found in Table \ref{table:Results}.  

The first column displays the names of the properties that were instrumented and compiled inside the Android operating system. The second and third columns, respectively, state the average time taken in milliseconds for the un-instrumented and instrumented events to finish. The last column shows the performance overhead introduced by the corresponding instrumented properties as a percentage of the original execution. 

\begin{table}[t]
\begin{center}
    \caption{The execution time for monitored and unmonitored action sequences}
    \label{table:Results}
    \begin{tabular}{ | l || r | r | r |}
    \hline
    Property & Time (ms) & Time (ms)   & Overhead \\ 
	name     & unmonitored & monitored & (\%) \\ \hline\hline
    WiFiLmt & 47.5 & 48.4 & 1.89\% \\ \hline
    PhoneExtBlk & 55.0 & 58.6 & 6.49\% \\ \hline
    PhoneTimeLim & 55.0 & 59.9 & 8.93\% \\ \hline
    MsgTimeLmt & 87.4 & 108.2 & 23.82\% \\ \hline
    MsgCntLmt & 87.4 & 109.2 & 24.96\% \\ \hline
    URLBlkReq & 11.9 & 14.9 & 25.31\% \\ \hline
    MsgCntLmtHr & 87.4 & 113.0 & 29.34\% \\ \hline
    FileAccessLmt & 0.04 & 0.1 & 34.65\% \\ \hline
    \end{tabular}
\end{center}
\end{table}

The properties introduced to the operating system resulted in runtime overheads between 1.89\% and 34.65\%. 
The difference in the monitoring complexity of the  properties can also be seen to affect the performance overheads, for instance when taking into consideration the \textit{PhoneTimeLim} and \textit{PhoneExtBlk} --- both monitor the same basic event but have different conditions. Although the highest performance overhead calculated was a considerable 34.65\%, taking 0.15 milliseconds longer to execute, the events of interest occur sparsely during normal use and in fact no delays were noticeable from a user perspective. 


\subsubsection{Measuring Memory usage}

Given that our approach offers the ability for application and global side monitors to store information, the impact on memory is another concern. None of the properties we specified for the case studies required any significant amount of memory, and the only property inducing a noticeable overhead in terms of memory was \emph{MsgCntLmtHr} (since it required keeping track of the timestamp of the previous 99 messages sent), which was still negligible. 

To confirm our intuition we also measured memory overheads whilst using the device. We note that since our monitoring framework affects the mobile device as a whole, the results shown in Table~\ref{table:ResultsMem} reflect the overall memory usage rather than those for any particular application. 
We note that the empirical observations confirm our intuition that the overall memory usage for both experiments is almost identical and the small recorded fluctuations could in reality have been caused by other factors. 

\begin{table}[h!]
\begin{center}
    \caption{Overall memory usage for both monitored and unmonitored devices with different applications}
    \label{table:ResultsMem}
    \begin{tabular}{ | l || c | c |}
    \hline
    Mobile  & Memory (MB) & Memory (MB)\\ 
    state & unmonitored & monitored \\ \hline \hline
    idle & 157 & 157 \\ \hline
    using phone app & 183 & 183 \\ \hline
    using messaging app & 184 & 185 \\ \hline
    using 1024 game app & 194 & 194 \\ \hline
    using puzzle app & 222 & 223 \\ \hline
    \end{tabular}
\end{center}
\end{table}


%% file: related-work.tex
\section{Related Work}
\label{sec:relatedWork}

There is a significant number of runtime verification approaches that aim to observe and in some cases even control application and user behaviour. Although most of these approaches allow identifying which events to monitor, where and how to instrument the monitors and how to act upon an observed behaviour, a common limitation is that most of the existing work allows only for application-centric properties. Our framework is capable of observing both application and device-centric properties. 


In \cite{AndSec,TaintDroid,DroidMat} the identification of malicious behaviour in third party applications is achieved by introducing loadable kernel modules inside the OS kernel that are able to observe system calls e.g. \textit{Read} and \textit{Write}. This approach was not used to observe events since such system calls are typically too low level and can, in some cases, vary in behaviour across devices. In addition, such a more fine-grained approach would mean that controlling the behaviour of the device can be more challenging, since additional calls would need to be modified.

Other work such as \cite{WeaveDroid,RVDroid,MTLAndroid,AppTrace} focuses on monitoring application behaviour at a much higher level of abstraction. Instead of observing system calls, such techniques focus on monitoring Java and Android framework API calls that are much easier to understand and have more attainable information that might be useful. Two popular approaches that are used to instrument monitors to observe such events include: replacing the stock application installer with one that automatically instruments monitors inside applications and directly instrumenting Android's middleware. 

Instrumenting monitors inside apps through the use of a custom installer is not always possible as discussed in \cite{WeaveDroid,RVDroid} due to limited memory and performance capabilities. To circumvent this limitation, a cloud-based approach is available. We did not consider this approach since we did not want to introduce unnecessary instrumented overheads when installing apps on the device. 

In our case, for every application that is executed, an instrumented instance of the Android runtime is initialized. This makes part of the Android middleware, hence making it an ideal location for monitors to be instrumented. An advantage over the custom installer regards the fact that the monitor instrumentation is only carried out once and does not suffer from performance or memory limitations. Our framework takes advantage of this approach and instruments the Android and Java frameworks located in the middle-ware to be able to observe system and application behaviour at will with the added benefit of being more difficult to tamper with. 

%% file: conclusions.tex
\section{Conclusions and future work}
\label{sec:conclusions}

The work we have presented, attempts to provide a novel runtime verification framework capable of monitoring device-centric properties requiring a global view to span multiple applications. 
The solution took into consideration the fact that most mobile device operating systems segregate applications, having a central manager for resource utilization, communication between applications and so on. 
In this context, the proposed monitoring architecture introduces monitors both at application and global level to capture events of interest and verify behaviour. This arrangement also allows monitors to communicate with each other through shared data, and to enjoy privileges which enable them to take corrective actions. 

The technique was implemented by introducing device-centric monitors; on the application side by injecting them inside the Android and Java API frameworks which are used by every running application; and globally by inserting a monitor at the kernel level. 
This approach also had to identify how the global monitor could communicate with the application monitor since the conventional approach of communicating monitors could not be na\"{i}vely implemented due to the possible lack of permissions allocated to the applications being monitored. 

To differentiate between global and application-side monitors, we have also presented a specification language enabling the user to tag which monitoring components belong where. Finally, we evaluated our approach on two typical case studies and found the specification to be adequately expressive and that the resulting overheads do not hinder the usability of the mobile device. 

\textit{Future work:} 
At this point, the implemented instrumentation is not fully automated, requiring the user to specify the exact entry and exit points of the event source code. In the future we plan to insert monitors directly inside the Dalvik/Java bytecode of the instrumented frameworks, 
	greatly decreasing the instrumentation time required, since the operating system would not need to be re-compiled and installed. 

Another limitation of the current implementation is that the state of the monitors are being kept in memory which is volatile. Therefore as soon as the instrumented device is restarted, all the monitors loose their state. We envisage an approach enabling us to save and restore the monitor state ensuring that the monitors resume where they left off after a restart. This approach would be implemented as an added monitoring module without the need to further modify the operating system. 


In the future we also hope to improve upon the communication between application and global side monitors and to save the specification writer the task of tagging monitoring subparts as application/global side through the use of static analysis.